\documentclass[letterpaper]{article} % DO NOT CHANGE THIS
\usepackage{aaai25}  % DO NOT CHANGE THIS
\usepackage{times}  % DO NOT CHANGE THIS
\usepackage{helv et}  % DO NOT CHANGE THIS
\usepackage{courier}  % DO NOT CHANGE THIS
\usepackage[hyphens]{url}  % DO NOT CHANGE THIS
\usepackage{graphicx} % DO NOT CHANGE THIS
\usepackage{natbib}  % DO NOT CHANGE THIS AND DO NOT ADD ANY OPTIONS TO IT
\usepackage{caption} % DO NOT CHANGE THIS AND DO NOT ADD ANY OPTIONS TO IT
\usepackage{algorithm}
\usepackage{amsmath}
\usepackage{booktabs}
\usepackage{multirow}
\usepackage{subfigure}
\usepackage{amssymb}
\usepackage{algpseudocode}

\usepackage{newfloat}
\usepackage{listings}
\DeclareCaptionStyle{ruled}{labelfont=normalfont,labelsep=colon,strut=off} % DO NOT CHANGE THIS
\floatstyle{ruled}
\newfloat{listing}{tb}{lst}{}
\floatname{listing}{Listing}
\title{MIETT: Multi-Instance Encrypted Traffic Transformer \\for Encrypted Traffic Classification}
\author {
    Xu-Yang Chen\equalcontrib,
    Lu Han\equalcontrib,
    De-Chuan Zhan,
    Han-Jia Ye\thanks{Corresponding Author}
}
\affiliations {
    School of Artificial Intelligence, Nanjing University, China \\
   National Key Laboratory for Novel Software Technology, Nanjing University, China\\
    \texttt{\{chenxy, hanlu, zhandc, yehj\}@lamda.nju.edu.cn}
}
\usepackage{bibentry}
\begin{document}

\maketitle
\begin{abstract}
% Encrypted traffic classification is a crucial research area in network security.
% 需要添加任务描述以及数据描述
Network traffic includes data transmitted across a network, such as web browsing and file transfers, and is organized into packets (small units of data) and flows (sequences of packets exchanged between two endpoints). Classifying encrypted traffic is essential for detecting security threats and optimizing network management.
Recent advancements have highlighted the superiority of foundation models in this task, particularly for their ability to leverage large amounts of unlabeled data and demonstrate strong generalization to unseen data. 
However, existing methods that focus on token-level relationships fail to capture broader flow patterns, as tokens, defined as sequences of hexadecimal digits, typically carry limited semantic information in encrypted traffic. These flow patterns, which are crucial for traffic classification, arise from the interactions between packets within a flow, not just their internal structure.
To address this limitation, we propose a Multi-Instance Encrypted Traffic Transformer (MIETT), which adopts a multi-instance approach where each packet is treated as a distinct instance within a larger bag representing the entire flow. This enables the model to capture both token-level and packet-level relationships more effectively through Two-Level Attention (TLA) layers, improving the model's ability to learn complex packet dynamics and flow patterns.
We further enhance the model's understanding of temporal and flow-specific dynamics by introducing two novel pre-training tasks: Packet Relative Position Prediction (PRPP) and Flow Contrastive Learning (FCL).  After fine-tuning, MIETT achieves state-of-the-art (SOTA) results across five datasets, demonstrating its effectiveness in classifying encrypted traffic and understanding complex network behaviors. Code is available at \url{https://github.com/Secilia-Cxy/MIETT}.
\end{abstract}

% Uncomment the following to link to your code, datasets, an extended version or similar.
%
% \begin{links}
% %     \link{Code}{https://aaai.org/example/code}
% %     % \link{Datasets}{https://aaai.org/example/datasets}
%     \link{Extended version}{https://aaai.org/example/extended-version}
% \end{links}

\section{Introduction}

Network traffic refers to the flow of data transmitted between devices over a network, typically structured into packets, which are small units sent across the network, and flows, which are sequences of packets exchanged between two points. Each packet is composed of two parts: the header and the payload. The header contains essential information, such as routing details, source and destination addresses, and packet length, while the payload carries the actual data being transmitted, which may be encrypted for security purposes.

\begin{figure}[t]
    \centering
\includegraphics[width=1\linewidth]{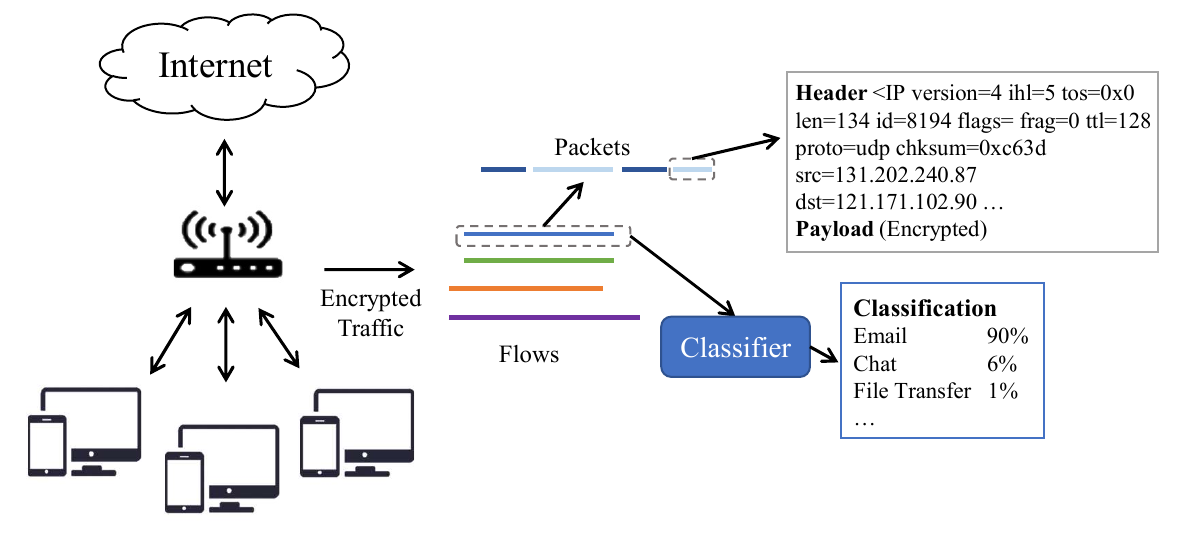}
    \caption{Encrypted traffic classification task description. Raw traffic is first divided into session flows, with each flow further segmented into a sequence of packets. A packet typically consists of a header and a payload. The task is to classify the type of a given flow.}
    \label{fig:task}
\end{figure}

Traffic classification, the process of identifying and categorizing network traffic, is crucial for both network management and cybersecurity. It allows network administrators to ensure Quality of Service (QoS), optimize bandwidth and detect malicious activities, thus maintaining network security and efficiency. The overview of encrypted traffic classification task is provided in Figure~\ref{fig:task}.

However, the increasing prevalence of encryption has made traditional classification methods, such as port-based and statistics-based approaches, less effective.
The advent of deep learning (DL) brought significant improvements, with payload-based methods using models like CNNs to automatically extract features from raw data. Despite their success, these methods rely heavily on large amounts of labeled data, which can be difficult to obtain.

Recently, foundation models have emerged as a powerful alternative, pre-trained on large volumes of unlabeled data and fine-tuned for specific tasks. PERT \cite{PERT} employs a BERT-based model with a Masked Language Modeling (MLM) task to pre-train a packet-level encoder, but this approach primarily focuses on token-level relationships within individual packets, overlooking the broader context of inter-packet relationships. ET-BERT \cite{LinXGLSY22} addresses this limitation by introducing the Same-origin BURST Prediction (SBP) task, which determines whether one packet follows another within the same flow. While this method accounts for adjacent packet relationships, it still falls short of capturing the full complexity of flow-level interactions and the broader context across multiple packet segments. YaTC \cite{zhaoyatc23} takes a different approach by tokenizing traffic data into patches and using a Masked Auto-Encoder (MAE) for pre-training a token-level encoder within a flow.
% However, focusing solely on token dependencies is less effective due to the low semantic information in encrypted traffic tokens. Instead, learning packet representations that reflect packet patterns proves to be more robust.
However, focusing solely on token dependencies is less effective due to the low semantic information in encrypted traffic tokens. This approach tends to overlook the broader patterns and relationships that exist between entire packets within a traffic flow. Instead, learning packet representations that capture these packet patterns proves to be more robust. 
% 这里缺一些观察实验，表示这两个任务分别有什么作用
Given that each packet can be viewed as a distinct instance carrying unique information within a flow, it is crucial to effectively model the relationships between packets to achieve a comprehensive flow representation. To address this, we propose a Multi-Instance Encrypted Traffic Transformer (MIETT) with Two-Level Attention (TLA) layers that capture both token-level and packet-level relationships. To further enhance the model's ability to capture temporal and flow-specific dynamics, we introduce two novel pre-training tasks: Packet Relative Position Prediction (PRPP) and Flow Contrastive Learning (FCL). The PRPP task is designed to help the model understand the sequential relationships between packets within a flow by predicting their relative positions, thereby enabling a more accurate representation of the flow’s structure. Meanwhile, the FCL task focuses on distinguishing packets from the same flow and those from different flows by learning robust representations that emphasize intra-flow similarities and inter-flow differences.

Building on the robust representations learned during pre-training, the model is then fully fine-tuned on the specific classification task. During this stage, both the packet encoder and the flow encoder are jointly optimized to adapt the model to the task at hand. 
% This full fine-tuning ensures that the model effectively leverages the representations learned during pre-training to achieve high performance in traffic flow classification.
Across experiments conducted on five datasets, MIETT consistently demonstrates competitive or superior performance compared to existing methods.
In conclusion, this paper presents several key contributions to the field of encrypted traffic classification:
\begin{itemize}
    \item We propose a novel Multi-Instance Encrypted Traffic Transformer (MIETT) architecture, which introduces Two-Level Attention (TLA) layers to effectively capture both token-level and packet-level relationships within traffic flows. 
    % This architecture is designed to handle the complexity of encrypted traffic by modeling the relationships between packets as distinct instances, thus providing a more comprehensive flow representation.
    \item We introduce two innovative pre-training tasks, Packet Relative Position Prediction (PRPP) and Flow Contrastive Learning (FCL). The PRPP task enhances the model's understanding of the sequential order of packets within a flow, while the FCL task improves the model's ability to differentiate between packets from the same flow and those from different flows. 
    % These tasks are crucial in capturing the temporal and flow-specific dynamics necessary for accurate traffic classification.
    \item We provide extensive empirical validation of the proposed MIETT model on multiple datasets, showing that our approach outperforms existing methods in terms of accuracy and F1-score.
\end{itemize}

\section{Related work}
% \subsection{Encrypted Traffic Classification}
Encrypted traffic classification is an essential research area in network security. As encryption becomes more prevalent, traditional traffic analysis methods that inspect packet payloads have become less effective. Researchers now focus on methods to classify encrypted traffic without accessing its content directly. These methods can be divided into three types: port-based methods, statistics-based methods, and payload-based methods.

\paragraph{Port-based Methods.}
Port-based traffic classification, one of the oldest methods, relies on associating port numbers in TCP/UDP headers with well-known IANA port numbers. Although it is fast and simple, the method's accuracy has declined due to port obfuscation, dynamic port assignments, network address translation (NAT), and other techniques~\cite{moore2005toward}.

\paragraph{Statistics-based Methods.}
Statistics-based methods for encrypted traffic classification leverage features independent of payloads, such as packet sizes, timing, and flow duration, to analyze and categorize traffic. Wang et al. computed entropy from packet payloads to classify eight traffic classes, using an SVM algorithm to select features~\cite{wang2011using}. Similarly, Korczynski and Duda focused on packet sizes, timing, and communication patterns for classifying traffic in encrypted tunnels~\cite{korczynski2012classifying}. However, these methods are based on manually designed features and are now outdated due to advancements in encryption protocols, changes in traffic patterns, and the use of traffic obfuscation techniques.

\paragraph{Payload-based DL Methods.} 
Payload-based methods usually leverage deep learning to analyze raw packet data, eliminating the need for manually designed features. These approaches have significantly advanced encrypted traffic classification by automatically extracting discriminative features. Deep Packet~\cite{lotfollahi2020deep} employs a CNN within a framework that includes a stacked autoencoder and convolutional neural network to classify network traffic. TSCRNN~\cite{lin2021tscrnn} utilizes CNN to extract abstract spatial features and introduces a stacked bidirectional LSTM to learn temporal characteristics. BiLSTM\_ATTN~\cite{yao2019identification} combines attention mechanisms with LSTM networks for enhanced encrypted traffic classification. Though these methods can automatically extract features, they heavily rely on labeled data, which is often difficult to obtain in large quantities for training.

\paragraph{Payload-based Foundation Models.} 
In recent years, pre-trained foundation models have become popular for addressing this issue. These models are pre-trained on large amounts of unlabeled data and fine-tuned on downstream labeled tasks. PERT~\cite{PERT} and ET-BERT~\cite{LinXGLSY22} tokenize traffic data using a vocabulary. PERT employs a Masked Language Model (MLM) pre-training task, while ET-BERT utilizes a modified MLM and Next Sentence Prediction (NSP) task. Additionally, YaTC~\cite{zhaoyatc23}, a vision model adaptation, tokenizes traffic data into patches and uses a Mask Auto-Encoder (MAE) for pre-training. However, these approaches do not adequately consider the unique structure of traffic flows and the relationships between packets. 
% We extend the potential of ET-BERT by designing the Multi-Instance Encrypted Traffic Transformer architecture and utilizing novel PRPP and FCL tasks. 
To address these limitations, we introduce the Multi-Instance Encrypted Traffic Transformer (MIETT) architecture, which leverages novel Packet Relative Position Prediction (PRPP) and Flow Contrastive Learning (FCL) tasks to better capture the complexities of traffic flows.

\section{Multi-Instance Encrypted Traffic Transformer}
In the task of encrypted traffic classification, we are provided with raw network traffic (PCAP traces) as input, and the objective is to classify it into categories, such as VPN services (e.g., P2P, streaming, email) and applications. In this section, we first outline the preprocessing steps that transform the raw data into a multi-instance traffic representation. We then introduce our Multi-Instance Encrypted Traffic Transformer (MIETT) architecture, which is specifically designed to handle and classify this traffic data efficiently. 

\subsection{MIETT Encoder}
This section details the process of representing multi-instance traffic data for use in the Multi-Instance Encrypted Traffic Transformer (MIETT) architecture. The process involves three key steps: tokenization of the raw data, representation of individual packets, and the aggregation of these packet representations into a unified flow representation.
\paragraph{Tokenization.}
\begin{figure}[t]
    \centering
    \includegraphics[width=0.6\linewidth]{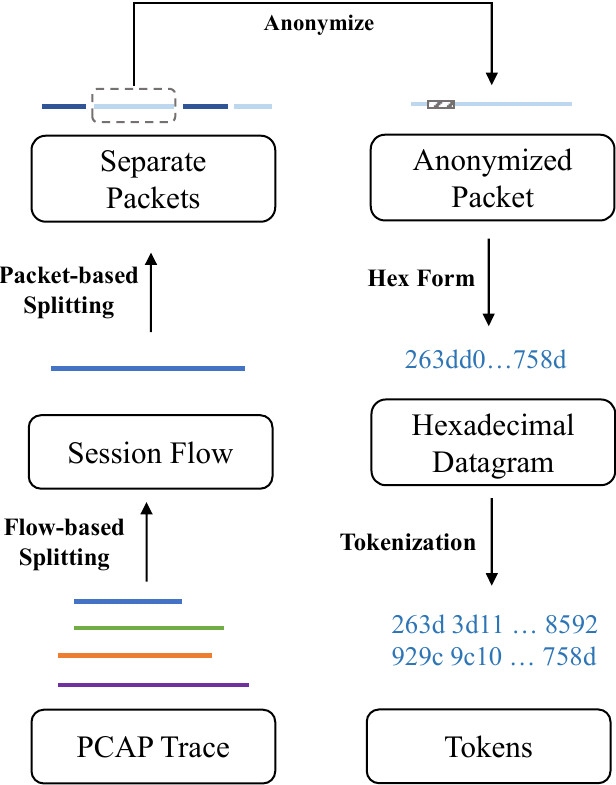}
    \caption{Data preprocessing. The raw traffic (PCAP trace) is first split into session flows and then further divided into individual packets. To protect data privacy, each packet is anonymized by masking the source and destination IP addresses and port numbers (replacing them with 0). The packet is then converted to its hexadecimal form, which is tokenized using a bi-gram model.}
    \label{fig:tokenization}
\end{figure}
The hexadecimal sequence of the flow is obtained through the data preprocessing steps outlined in Figure~\ref{fig:tokenization}.
Following ET-BERT~\cite{LinXGLSY22}, we encode the hexadecimal sequence using a bi-gram model, where each unit consists of two consecutive bytes. We then utilize Byte-Pair Encoding (BPE) for token representation, with token units ranging from 0 to 65535 and a maximum dictionary size of 65536. For training, we also incorporate special tokens, including [CLS], [PAD], and [MASK]. 

\paragraph{Packet Representation.}
Each packet begins with a [CLS] token, followed by tokens extracted from the packet's content, which includes a header containing meta-features and an encrypted payload. The embedding of each token is obtained by combining two parts: position embedding, which indicates the token's position within the packet, and value embedding. To ensure efficient information utilization and avoid excessive focus on long packets, the packet length is standardized to a fixed size of 128. Packets shorter than 128 tokens are padded with [PAD] tokens at the end.

\paragraph{Flow Representation.}
The flow representation consists of multiple packet representations, which are stacked to form a matrix $\mathbf{X} \in \mathbb{R}^{N \times L \times d}$, where $N$ is the number of packets, $L$ is the packet length, and $d$ is the embedding dimension. This multi-instance representation, as opposed to the previous method employed by ET-BERT of directly concatenating packets, allows for more effective modeling of the relationships between packets and better captures the organizational structure of flows. 

\subsection{MIETT Architecture}
The flow representation, a 2D map of tokens, can be flattened into a 1D sequence for input into a standard transformer, as done in ViT~\cite{alexViT21} and YaTC~\cite{zhaoyatc23} during pre-training and fine-tuning. However, this approach introduces two issues: (1) Flattening the sequence loses temporal information, such as packet order, potentially overlooking important dependencies. While 2D position embeddings could indicate temporal relationships, they are inflexible and may fail in scenarios like packet loss. (2) The computational complexity, which is $O(N^2L^2d)$, increases significantly with the inclusion of more packets due to the extended sequence length. To address these concerns, we introduce the Two-Level Attention (TLA) layer to preserve temporal structure and maintain computational efficiency.

\paragraph{Overall Architecture.}

\begin{figure*}[t]
    \includegraphics[width=\linewidth]{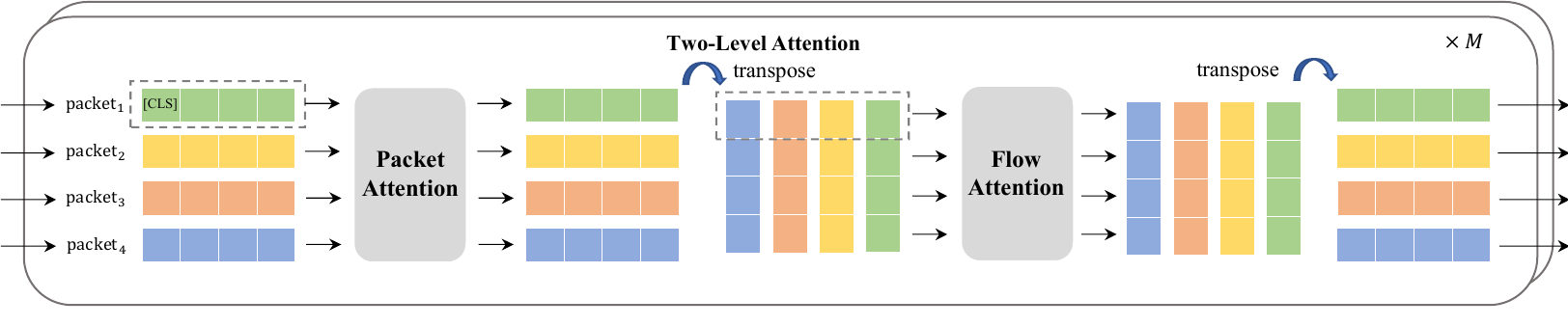}
    \caption{Overall architecture of the Multi-Instance Encrypted Traffic Transformer (MIETT). After passing through the MIETT encoder, the flow representation is processed by $M$ Two-Level Attention (TLA) layers, each comprising a packet attention mechanism and a flow attention mechanism.}
    \label{fig:method}
\end{figure*}

The Multi-Instance Encrypted Traffic Transformer (MIETT) begins by embedding the tokens as described in the MIETT Encoder section. Following this, the flow representation is processed through $M$ TLA layers. Finally, the embeddings of the [CLS] tokens from all the packets are utilized for pre-training or fine-tuning tasks. The overall architecture of MIETT can be found in Figure~\ref{fig:method}.

\paragraph{Two-Level Attention (TLA) Layer.}
The TLA layer captures both intra-packet and inter-packet dependencies to enhance the model's understanding of complex traffic flows. It operates in two stages: Packet Attention and Flow Attention. 

In the Packet Attention stage, Multi-Head Self-Attention (MHSA) is applied within individual packets to identify dependencies between tokens, ensuring the model understands the internal structure of each packet. For MHSA, the key, query, and value are all set to the input sequence, with the output being an enriched sequence representation.

In the Flow Attention stage, MHSA is used across packet representations at each position to capture dependencies between tokens across different packets. This two-stage approach effectively models the hierarchical structure of traffic flows by combining detailed packet-level insights with broader inter-packet relationships.

This method significantly improves the model's ability to capture complex dependencies but also introduces computational complexity. The MHSA in Packet Attention has a complexity of $O(L^2 d)$ per packet, while Flow Attention has a complexity of $O(N^2 d)$ per position, resulting in an overall complexity of $O(N L^2 d + L N^2 d)$. For default values of $L=128$ and $N=5$, our method is approximately 4.8 times more efficient than flattening the sequence to 1D and feeding it into a standard transformer.

\paragraph{Packet Attention.} 
In the first stage of the TLA layer, we focus on the intra-packet relationships by performing Multi-Head Self-Attention (MHSA) within individual packets. Let $\mathbf{X} \in \mathbb{R}^{N \times L \times d}$ be the flow representation, where $N$ is the number of packets, $L$ is the packet length, and $d$ is the embedding dimension.

For each packet $\mathbf{X}_i \in \mathbb{R}^{L \times d}$, we apply self-attention to capture the dependencies between the tokens within the packet. The process is as follows:
\begin{align}
    \hat{\mathbf{X}}_i^{\text{pkt}} & = \textbf{LayerNorm}(\mathbf{X}_i + \textbf{MHSA}^{\text{pkt}}(\mathbf{X}_i)) \\
% \end{equation}
% \begin{equation}
    \mathbf{X}_i^{\text{pkt}} &= \textbf{LayerNorm}(\hat{\mathbf{X}}_i^{\text{pkt}} + \textbf{MLP}(\hat{\mathbf{X}}_i^{\text{pkt}}))
\end{align}

\paragraph{Flow Attention.}
In the second stage of the TLA layer, we focus on the inter-packet relationships by performing multi-head self-attention (MHSA) across the packet representations at each position within the packets. Let $\mathbf{X}^{packet} \in \mathbb{R}^{N \times L \times d}$ be the updated flow representation after the packet attention stage.

For each position $j$ (where $j \in {1, 2, \ldots, L}$) within the packets, we gather the token representations across all packets, resulting in a matrix $\mathbf{X}^{packet}_{\cdot j} \in \mathbb{R}^{N \times d}$, where $N$ is the number of packets. We apply self-attention to these matrices to capture the dependencies between tokens across different packets. The process is as follows:

\begin{align}
\hat{\mathbf{X}}^{\text{flow}}_{\cdot j} &= \textbf{LayerNorm}(\mathbf{X}^{\text{pkt}}_{\cdot j} + \textbf{MHSA}^{\text{flow}}(\mathbf{X}^{\text{pkt}}_{\cdot j})) \\
\mathbf{X}^{\text{flow}}_{\cdot j} &= \textbf{LayerNorm}(\hat{\mathbf{X}}^{\text{flow}}_{\cdot j} + \textbf{MLP}(\hat{\mathbf{X}}^{\text{flow}}_{\cdot j}))
\end{align}

\section{Training Tasks}

This section outlines the training tasks designed to improve the model's ability to classify encrypted network traffic. The pre-training phase includes three tasks: Masked Flow Prediction (MFP), Packet Relative Position Prediction (PRPP), and Flow Contrastive Learning (FCL). These tasks help the model capture flow dependencies, predict packet order, and differentiate flow-level features. After pre-training, the model is fine-tuned for traffic flow classification, optimizing it for the final task.
\subsection{Pre-Training tasks}

\begin{figure*}[t]
    \includegraphics[width=\linewidth]{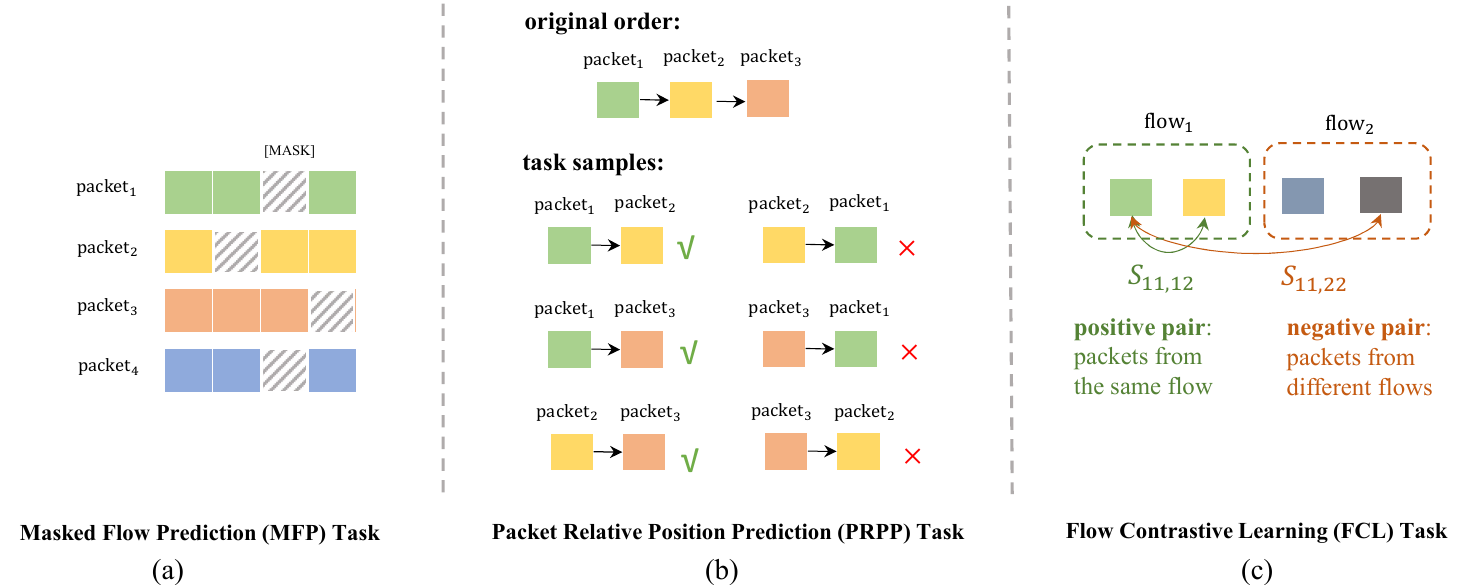}
    \caption{Overview of pre-training tasks. (a) Masked Flow Prediction (MFP) Task: The model is tasked with predicting the original content of masked tokens using the context provided by the unmasked tokens. (b) Packet Relative Position Prediction (PRPP) Task: The model's objective is to determine, for each pair of packets (i, j), whether packet i precedes packet j. (c) Flow Contrastive Learning (FCL) Task: The goal is to ensure that packets within the same flow (positive pairs) are more similar in the embedding space, while packets from different flows (negative pairs) are less similar.}
    \label{fig:pretrain}
\end{figure*}

The model we propose consists of several Two-Level Attention (TLA) layers, designed to effectively capture both packet-level and flow-level information within traffic flows. During the pre-training stage, We utilize a pre-trained ET-BERT checkpoint for the packet attention, which is kept frozen during training, while the flow attention is trained to learn the overall structure and dependencies within the flows. This design allows our model to leverage established packet-level features while focusing on improving flow-level understanding. The general view of our proposed 3 pre-training tasks can be found in Figure~\ref{fig:pretrain}.

\paragraph{Masked Flow Prediction (MFP) Task.}
The Masked Flow Prediction (MFP) task is aimed at enhancing the model’s ability to handle incomplete information within a traffic flow. In this task, 15\% of the tokens within a flow are randomly masked, and the model is tasked with predicting the original content of these masked tokens using the context provided by the unmasked tokens. By training the flow encoder to infer the missing tokens, the model learns to capture the underlying structure and dependencies within the traffic flow. 

\paragraph{Packet Relative Position Prediction (PRPP) Task.}
The Packet Relative Position Prediction (PRPP) task is designed to predict the relative order of packets within a flow, based on the embeddings of the [CLS] tokens extracted from each packet. This task helps the model understand the sequential relationships between packets, which is critical for accurately modeling traffic flows.

Given an output flow representation with $N$ packets, let $\mathbf{O}^{\text{pkt}} \in \mathbb{R}^{N \times d}$ represent the embeddings of the [CLS] tokens from all packets in the flow, where $d$ is the embedding dimension. The task is to determine, for each pair of packets $(i, j)$, whether packet $i$ comes before packet $j$.

Firstly, a linear transformation is applied followed by an activation function and layer normalization to [CLS] tokens:
\begin{equation}
    \mathbf{P} = \textbf{LayerNorm}(\textbf{GELU}( \mathbf{O}^{\text{pkt}}\mathbf{W}_1 + \mathbf{b}_1))
\end{equation}
where $\mathbf{W}_1 \in \mathbb{R}^{d \times d}$ and $\mathbf{b}_1 \in \mathbb{R}^d$ are learnable parameters. $\mathbf{P} \in \mathbb{R}^{N \times d}$.

Then, the predicted relative position of packets $(i, j)$ is computed as follows:
\begin{equation}
    \hat{z}_{ij} = \textbf{Softmax}((\mathbf{P}_i - \mathbf{P}_j)\mathbf{W}_2  + \mathbf{b}_2)
\end{equation}
where $\mathbf{W}_2 \in \mathbb{R}^{d \times 2}$ and $\mathbf{b}_2 \in \mathbb{R}^2$ are learnable parameters, and $\hat{z}_{ij} \in \mathbb{R}^2$ gives the probability that packet $i$ comes before or after packet $j$.

The ground truth labels $z_{ij} \in \{0, 1\}$ are based on the original order of packets:
\begin{equation}
    z_{ij} = 
\begin{cases} 
1 & \text{if packet } i \text{ comes before packet } j \\
0 & \text{otherwise}
\end{cases}
\end{equation}

Finally, the PRPP loss $\mathcal{L}_{\text{PRPP}}$ is calculated using cross-entropy between the predicted labels and the ground truth:
\begin{equation}
    \mathcal{L}_{\text{PRPP}} = -\sum_{i,j,i \neq j} z_{ij} \log(\hat{z}_{ij}) + (1 - z_{ij}) \log(1 - \hat{z}_{ij})
\end{equation}

\paragraph{Flow Contrastive Learning (FCL) Task.}
The Flow Contrastive Learning (FCL) task enhances the model's ability to differentiate between traffic flows by learning robust representations. The objective is to ensure that packets within the same flow (positive pairs) are more similar in the embedding space, while packets from different flows (negative pairs) are less similar. Notably, both positive and negative pairs are constructed using identical packet positions within their respective flows, maintaining consistency in the comparisons.

% Given several output flow representations within a batch, let $\mathbf{O}^{\text{flow}} \in \mathbb{R}^{BS \times N \times d}$ represent the embeddings of the [CLS] tokens from all packets in the flows in a batch, where $BS$ is the batch size, $N$ is the number of packets per flow, and $d$ is the embedding dimension. 

Given a batch of output flow representations, let $\mathbf{O}^{\text{flow}} \in \mathbb{R}^{BS \times N \times d}$ represent the embeddings of the [CLS] tokens from all packets in the flows within the batch, where $BS$ denotes the batch size, $N$ the number of packets per flow, and $d$ the embedding dimension.

First, a multi-layer perceptron (MLP) is applied to each [CLS] token:
\begin{align}
    \mathbf{C} &= \textbf{LayerNorm}(\textbf{GELU}( \mathbf{O}^{\text{flow}}\mathbf{W}_3 + \mathbf{b}_3)) \\
    \mathbf{C} &=  \mathbf{C}\mathbf{W}_4 + \mathbf{b}_4
\end{align}
where $\mathbf{W}_3 \in \mathbb{R}^{d \times d}$, $\mathbf{b}_3 \in \mathbb{R}^d$, $\mathbf{W}_4 \in \mathbb{R}^{d \times d}$, and $\mathbf{b}_4 \in \mathbb{R}^d$ are learnable parameters. $\mathbf{C} \in \mathbb{R}^{BS \times N \times d}$.

\begin{table*}[ht]
\centering
  \setlength{\tabcolsep}{1.25mm} % Adjust the column separation
  % \resizebox{2.1\columnwidth}{!}{
\begin{tabular}{lcccccccccc}
\toprule
\multirow{2}{*}{\textbf{Methods}}         & \multicolumn{2}{c}{ISCXVPN 2016} & \multicolumn{2}{c}{ISCXTor 2016} & \multicolumn{2}{c}{CrossPlatform (Android)} & \multicolumn{2}{c}{CrossPlatform (iOS)} & \multicolumn{2}{c}{CICIoT 2023}  \\
\cmidrule(lr){2-3} \cmidrule(lr){4-5}\cmidrule(lr){6-7} \cmidrule(lr){8-9}\cmidrule(lr){10-11}
         & AC      & F1                     & AC      & F1                     & AC      & F1                                   & AC      & F1                             & AC      & F1                      \\
\midrule
Datanet         & 69.18\% & 13.63\%                & 49.81\% & 9.50\%                 & 9.45\%  & 1.53\%                               & 4.81\%  & 0.05\%                         & 2.50\%  & 0.81\%                  \\
Fs-Net          & 29.30\% & 33.67\%                & 82.03\% & 63.54\%                & 7.08\%  & 4.11\%                               & 10.94\% & 6.38\%                         & 66.80\% & 54.81\%                 \\
BiLSTM\_ATTN    & 0.57\%  & 3.13\%                 & 88.33\% & 55.54\%                & 0.45\%  & 0.04\%                               & 0.29\%  & 0.01\%                         & 5.79\%  & 4.66\%                  \\
DeepPacket      & 69.18\% & 13.63\%                & 49.81\% & 9.50\%                 & 4.84\%  & 0.04\%                               & 4.81\%  & 0.05\%                         & 34.35\% & 8.52\%                  \\
TSCRNN~         & 69.18\% & 13.63\%                & 44.70\% & 13.40\%                & 2.43\%  & 0.24\%                               & 2.84\%  & 0.51\%                         & 2.50\%  & 0.81\%                  \\
\midrule
YaTC            & \textbf{78.05\%} & 70.83\%                & \textbf{97.39\%} & \textbf{85.12}\%                & \underline{91.61\%} & \underline{82.28\%}                              & 75.31\% & 69.57\%                        & 86.18\% & 73.16\%                 \\
ET-BERT         & 74.62\% & \underline{71.10\%}                & 95.71\% & 80.29\%                & 84.63\% & 67.70\%                              & \underline{77.05\%} & \underline{74.26\%}                       & \underline{88.09\%} & \textbf{83.29\%}                 \\
\midrule
MIETT (ours) & \underline{76.07\%} & \textbf{77.86\% }                & \underline{96.60\%} & \underline{82.15\%}                & \textbf{93.00\%} & \textbf{82.36\%}                              & \textbf{79.63\%} & \textbf{75.03\%}                        & \textbf{88.53\%} & \underline{82.48\%} \\
\bottomrule
\end{tabular}
% }

\caption{Performance comparisons on encrypted traffic classification tasks. We denote the \textbf{best} and \underline{second-best} results with bold and underline.}
  \label{tab:main-results}
\end{table*}

Next, the similarity between two packets is computed using cosine similarity:
\begin{equation}
    \mathbf{S}_{i_1 j_1, i_2 j_2} = \frac{\mathbf{C}_{i_1 j_1}^T \mathbf{C}_{i_2 j_2}}{\|\mathbf{C}_{i_1 j_1}\| \|\mathbf{C}_{i_2 j_2}\|}
\end{equation}
where $i_1, i_2$ represent the flow IDs within the batch, and $j_1, j_2$ denote the packet positions within the flow. $\mathbf{C}_{i j}\in \mathbb{R}^{d}$.

Finally, the contrastive loss is computed using the similarity matrix $\mathbf{S}$:
\begin{equation}
    \mathcal{L}_{\text{FCL}} = - \sum_{\substack{i_1, j_1, j_2 \\ j_1\neq j_2}} \log\frac{\exp(\mathbf{S}_{i_1j_1, i_1j_2})}{\exp(\mathbf{S}_{i_1j_1, i_1j_2}) + \sum\limits_{i_2 \neq i_1} \exp(\mathbf{S}_{i_1j_1,i_2j_2}) }
\end{equation}
where $i_1, i_2 \in [1, BS]$ represent the flow IDs in the batch, $j_1, j_2 \in [1, N]$ denote the packet positions within the flow.

\paragraph{Conclusion.}
Overall, the final loss during the pre-train stage is the weighted sum of the above 3 losses:
\begin{equation}
    \mathcal{L}_{\text{pt}} = \mathcal{L}_{\text{MPF}} + \alpha \mathcal{L}_{\text{PRPP}} + \beta \mathcal{L}_{\text{FCL}}
\end{equation}
where $\alpha$ and $\beta$ are hyperparameters.

\subsection{Fine-Tuning Task}
% \begin{figure}[t]
%     \centering
%     \includegraphics[width=0.5\linewidth]{figure/finetune.pdf}
%     \caption{Output layer for fine-tuning task.}
%     \label{fig:ft}
% \end{figure}
The objective of the fine-tuning task is to classify a given flow into a specific class. After processing the flow through several TLA layers, we obtain a flow representation by extracting the embeddings of the [CLS] tokens from all packets. These embeddings, representing each packet, are then aggregated using mean pooling to form a comprehensive representation of the entire flow. This mean-pooled flow representation is passed through a multi-layer perceptron (MLP) to produce the final classification output, indicating the predicted class of the traffic flow. 

During this stage, the entire model, including both the packet encoder (previously frozen during pre-training) and the flow encoder, is fine-tuned. The fine-tuning process optimizes the model by minimizing the cross-entropy loss:

\begin{equation}
    \mathcal{L}_{\text{ft}} = - \sum_{c} y_c \log(\hat{y}_c)
\end{equation}
where $y_c$ is the true label, and $\hat{y}_c$ is the predicted probability for class $c$. 

\section{Experiments}
\subsection{Experiment Setup}

\paragraph{Datasets and Benchmarks.}

\begin{table}
\centering
\setlength{\tabcolsep}{0.5mm} % Adjust the column separation
% \resizebox{1\columnwidth}{!}{
\begin{tabular}{c|c|c|c}
\toprule
Dataset                                                  & \#Flow    & Task                        & \#Label \\
\midrule
ISCXVPN 2016                 & 311,390   & VPN Service       & 6       \\\midrule
ISCXTor 2016& 55,523    & Tor Service       & 7       \\\midrule
CrossPlatform (Android) & 66,346    & Application  & 212     \\\midrule
Cross Platform (iOS)  & 34,912    & Application  & 196     \\\midrule
CICIoT Dataset 2023  & 1,163,495 & IoT Attack & 7    \\
\bottomrule
\end{tabular}
% }
\caption{The statistical information of 5 different datasets.}
\label{tab:stats_of_dataset}
\end{table}

For the encrypted traffic classification task, we evaluate our method on five datasets: ISCXVPN 2016~\cite{draper2016iscxvpn}, ISCXTor 2016~\cite{lashkari2017iscxtor}, and the Cross-Platform~\cite{van2020crossplatform} dataset, which includes two subsets (Android and iOS), as well as the CIC IoT Dataset 2023~\cite{neto2023ciciot2023}. We utilize data preprocessed by Netbench~\cite{netbench}. The data used for pre-training consists of the training sets from all five datasets in Netbench, without labels. The dataset statistics and descriptions of the fine-tuning tasks are detailed in Table~\ref{tab:stats_of_dataset}. The data is split into training, validation, and test sets with a ratio of 8:1:1.

% \paragraph{Main Metrics.}
% The primary metrics for comparison are Accuracy and F1-Score. Accuracy measures the proportion of correct predictions, while F1-Score balances precision and recall.

% \begin{equation}
% \text{Accuracy} = \frac{TP + TN}{TP + TN + FP + FN}
% \end{equation}

% \begin{equation}
% \text{F1} = 2 \cdot \frac{\text{Precision} \cdot \text{Recall}}{\text{Precision} + \text{Recall}}
% \end{equation}

% \begin{equation}
% \text{Precision} = \frac{TP}{TP + FP}, \text{Recall} = \frac{TP}{TP + FN}
% \end{equation}

% Here, TP refers to True Positives, TN refers to True Negatives, FP refers to False Positives, and FN refers to False Negatives.

\paragraph{Compared Methods.}
We compare our method against 7 payload-based approaches, including deep learning methods such as Datanet~\cite{wang2018datanet}, Fs-Net~\cite{liu2019fs}, BiLSTM\_ATTN~\cite{yao2019identification}, DeepPacket~\cite{lotfollahi2020deep}, TSCRNN~\cite{lin2021tscrnn}, as well as foundation models like ET-BERT~\cite{LinXGLSY22} and YaTC~\cite{zhaoyatc23}.

\paragraph{Implementation Details.}
During the pre-training stage, we set the training steps to 150,000 and randomly select five of the first ten packets for training. The masking ratio for the Masked Flow Prediction (MFP) task is set to 15\%. The weights for the Packet Relative Position Prediction (PRPP) and MFP tasks are both set to 0.2. In the fine-tuning stage, we train for 30 epochs using the first five packets. For both stages, the packet length ($L$) is set to 128, the number of packets ($N$) is set to 5, the embedding dimension ($d$) is set to 768, and the number of Two-Level Attention (TLA) layers is set to 12. The learning rate is set to $2 \times 10^{-5}$, and the AdamW optimizer is used. All experiments are conducted on a server with two NVIDIA RTX A6000 GPUs.

\subsection{Main Results}
The primary metrics for comparison are Accuracy and F1-Score. Accuracy measures the proportion of correct predictions, while F1-Score balances precision and recall.
Table~\ref{tab:main-results} presents the performance comparison of various models on encrypted traffic classification tasks, with baseline results sourced from NetBench~\cite{netbench}. The results clearly indicate that traditional deep learning methods, such as DataNet, DeepPacket, FS-Net, TSCRNN, and BiLSTM\_ATTN, struggle to generalize effectively to new tasks, particularly on complex datasets. These models frequently exhibit a bias toward dominant classes, resulting in consistently low F1-scores, especially on datasets like CrossPlatform (Android) and CrossPlatform (iOS).

In contrast, our MIETT model demonstrates significant improvements in both accuracy and F1-scores across the board, showcasing its superior capability to handle the complexities of encrypted traffic. Notably, MIETT consistently achieves competitive or superior performance compared to existing methods. For instance, in the CrossPlatform (Android) dataset, MIETT outperforms ET-BERT with an 8.27\% increase in accuracy and a 14.66\% increase in F1-score, highlighting the effectiveness of the flow attention.

\subsection{Ablation Study}
\paragraph{Impact of Pre-Training Tasks.}
\begin{table}
\centering
\setlength{\tabcolsep}{0.3mm} % Adjust the column separation
% \resizebox{1\columnwidth}{!}{

\begin{tabular}{lcccc}
\toprule
  \multirow{2}{*}{\textbf{Methods}}  & \multicolumn{2}{c}{CrossPlatform
  (Android)} & \multicolumn{2}{c}{CrossPlatform (iOS)}  \\
  \cmidrule(lr){2-3} \cmidrule(lr){4-5}
                 & AC      & F1                                   & AC      & F1                              \\
\midrule
from scratch & 88.08\% & 73.62\%                              & 71.63\% & 63.43\%                         \\
w/o PRPP    & 90.79\% & 79.02\%                              & 78.96\% & 74.35\%                         \\
w/o FCL     & 91.90\% & 81.60\%                              & 79.46\% & 74.80\%                         \\
Ours  & \textbf{93.00\%} & \textbf{82.36\%}                              & \textbf{79.63\%} & \textbf{75.03\%}                   \\
\bottomrule
\end{tabular}
% }
\caption{Impact of Pre-Training Tasks.}
\label{tab:pre-train}
\end{table}

\begin{table}
\centering

\setlength{\tabcolsep}{0.3mm} % Adjust the column separation
% \resizebox{1\columnwidth}{!}{
\begin{tabular}{lcccc}
\toprule
  \multirow{2}{*}{\textbf{Methods}}  & \multicolumn{2}{c}{CrossPlatform
  (Android)} & \multicolumn{2}{c}{CrossPlatform (iOS)}  \\
  \cmidrule(lr){2-3} \cmidrule(lr){4-5}
                 & AC      & F1                                   & AC      & F1                              \\
\midrule
w/o pkt attn & 62.19\% & 28.59\%                              & 55.58\% & 39.93\%                         \\
w/o flow attn   & 91.85\% & 80.77\%                              & 79.11\% & 72.46\%                         \\
TLA (ours)  & \textbf{93.00\%} & \textbf{82.36\%}                              & \textbf{79.63\%} & \textbf{75.03\%}                   \\
\bottomrule
\end{tabular}
% }
\caption{Impact of TLA components, where 'pkt attn' refers to packet attention and 'flow attn' refers to flow attention.}
\label{tab:TLA-ablation}
\end{table}
Table~\ref{tab:pre-train} presents an ablation study that compares different versions of the pre-training tasks on the CrossPlatform (Android) and CrossPlatform (iOS) datasets to evaluate the contributions of specific tasks. The baseline model ("from scratch") shows decent performance, indicating that pre-training is important and can significantly enhance the model's ability to generalize to new tasks. As the results demonstrate, incorporating specific pre-training tasks such as Packet Relative Position Prediction (PRPP) and Flow Contrastive Learning (FCL) further boosts the model's performance. 

\paragraph{Impact of TLA Components.}

TLA captures intra- and inter-packet dependencies, efficiently handling the challenges of modeling token-to-token relations across packets. Table~\ref{tab:main-results} shows that ET-BERT, using only token-level attention, performs worse. As shown in Table~\ref{tab:TLA-ablation}, results on CrossPlatform(Android) shows removing flow attention raises error rates by 16.4\%, and using token-mean embeddings for packets lowers accuracy to 62.19\%, underscoring the importance of both attentions.
\paragraph{Impact of the Number of Packets.}
\begin{figure}[t]
    \centering
    \includegraphics[width=1\linewidth]{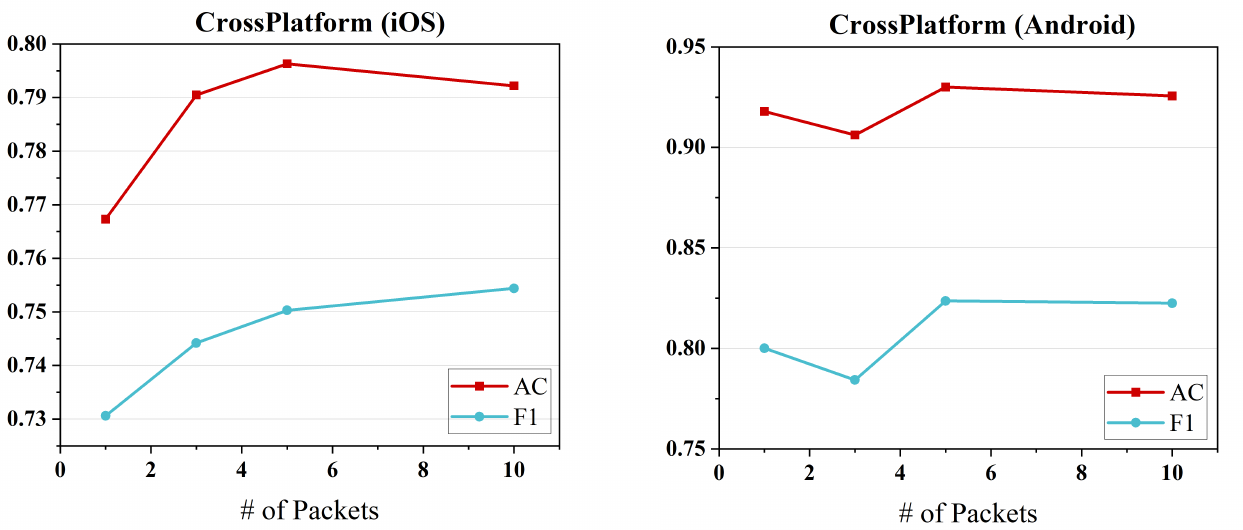}
    \caption{Impact of the Number of Packets.}
    \label{fig:pkt_num}
\end{figure}

Figure~\ref{fig:pkt_num} shows the impact of packet count. On the left, we observe that as more packets provide additional information, the F1 score increases, confirming our expectations. However, on the CrossPlatform (Android) dataset, using just one packet outperforms using three, with 91.79\% accuracy, higher than all baseline models using five packets. This suggests that in some datasets, the initial packets may contain the most critical information, and adding more packets may not always improve performance. If packet relationships are not effectively modeled, it could even harm performance.  In this case, the reason our model did not perform well may be due to the discrepancy between the pre-training stage, where five packets were used, and the fine-tuning stage, where only three were used, leading to a mismatch in distribution.

\paragraph{Impact of the Component of Packets.}
\begin{table}
\centering

\setlength{\tabcolsep}{0.3mm} % Adjust the column separation
% \resizebox{1\columnwidth}{!}{
\begin{tabular}{lcccc}
\toprule
  \multirow{2}{*}{\textbf{Methods}}  & \multicolumn{2}{c}{CrossPlatform
  (Android)} & \multicolumn{2}{c}{CrossPlatform (iOS)}  \\
  \cmidrule(lr){2-3} \cmidrule(lr){4-5}
                 & AC      & F1                                   & AC      & F1                              \\
\midrule
header only & 78.77\% & 64.98\%                              & 52.96\% & 41.66\%                         \\
payload only    & 72.46\% & 65.80\%                              & 63.82\% & 55.30\%                         \\
All  & \textbf{93.00\%} & \textbf{82.36\%}                              & \textbf{79.63\%} & \textbf{75.03\%}                   \\
\bottomrule
\end{tabular}
% }
\caption{Impact of the Component of Packets.}
\label{tab:component}
\end{table}

Table~\ref{tab:component} highlights the advantages of payload-based methods in traffic classification. When using only the packet header, the model's performance is significantly lower, showing that the header lacks sufficient information for accurate classification. However, when focusing on the payload, which contains the actual data, the model's F1-scores improve, especially on complex datasets like Cross Platform (iOS). This demonstrates that payload-based methods are more effective in capturing the essential characteristics of traffic, making them superior for encrypted traffic classification. Combining both header and payload yields the best results.
\section{Conclusion}
In this paper, we introduced MIETT to address challenges in encrypted traffic classification. The model uses TLA layers to capture both token-level and packet-level relationships. Through novel pre-training strategies, MIETT effectively learns temporal and flow-specific dynamics. Our experiments demonstrate that MIETT outperforms existing methods, achieving superior results across five datasets.

\section{Acknowledgments} 
This work is partially supported by the National Science and Technology Major Project (2022ZD0114805), NSFC (62376118), Key Program of Jiangsu Science Foundation (BK20243012), Collaborative Innovation Center of Novel Software Technology and Industrialization.

\bibliography{traffic}

\appendix
\section{Appendix}
We provide details omitted in the main paper. 

\subsection{Pseudo Code}
As described in the main section, we introduce a new architecture called the Multi-Instance Encrypted Traffic Transformer (MIETT), along with two novel pre-training tasks: the Packet Relative Position Prediction (PRPP) Task and the Flow Contrastive Learning (FCL) Task. Additionally, the fine-tuning task is presented. Below are their respective pseudocode implementations.
\begin{algorithm}[htp]
    \caption{MIETT Encoder}
    
    \begin{algorithmic}[1]
    \Require Raw traffic flow tokens $\mathbf{X}_{\text{raw}} \in \mathbb{R}^{B \times N \times L}$, where $B$ is the batch size, $N$ is the number of packets, and $L$ is the packet length. Position embedding function $\text{Position\_Emb}: \{0, 1, \dots, L-1\} \rightarrow \mathbb{R}^{d}$, which maps each position to a $d$-dimensional embedding. Value embedding function $\text{Value\_Emb}: \{0, 1, \dots, \text{Vocab\_size}-1\} \rightarrow \mathbb{R}^{d}$, which maps each token value to a $d$-dimensional embedding. 
            \State $\mathbf{X} = \text{Position\_Emb}(\mathbf{X}_{\text{raw}}) + \text{Value\_Emb}(\mathbf{X}_{\text{raw}})$
            
            \For{$l = 1\ldots M$}
            \State $\mathbf{X} = \textbf{LayerNorm}(\mathbf{X} + \textbf{MHSA}^{\text{pkt}}(\mathbf{X}))$ \Comment{Packet Attention, $\mathbf{X} \in \mathbb{R}^{B \times N \times L \times d}$}
            \State  $\mathbf{X} = \textbf{LayerNorm}(\mathbf{X} + \textbf{MLP}(\mathbf{X}))$
            \State $\mathbf{X} = \mathbf{X}^T$
            
            \State $\mathbf{X} = \textbf{LayerNorm}(\mathbf{X} + \textbf{MHSA}^{\text{flow}}(\mathbf{X}))$ \Comment{Flow Attention, $\mathbf{X} \in \mathbb{R}^{B \times L \times N \times d}$}
            \State  $\mathbf{X} = \textbf{LayerNorm}(\mathbf{X} + \textbf{MLP}(\mathbf{X}))$
            \State $\mathbf{X} = \mathbf{X}^T$
            \EndFor
            \State \textbf{Return} $\mathbf{X}$

    \end{algorithmic}
    \label{alg:model}
\end{algorithm}

\begin{algorithm}[htp]
    \caption{PRPP Task}
    \begin{algorithmic}[1]
        \Require $\mathbf{X} \in \mathbb{R}^{B \times N \times L \times d}$ is the output from the MIETT Encoder, where $B$ is the batch size, $N$ is the number of packets, $L$ is the packet length, and $d$ is the dimension of embedding. $\mathbf{W}_1 \in \mathbb{R}^{d \times d}, \mathbf{W}_2 \in \mathbb{R}^{d \times 2}$ and $\mathbf{b}_1 \in \mathbb{R}^d , \mathbf{b}_2 \in \mathbb{R}^2$ are learnable parameters. 
        \State $\mathbf{X}_\text{CLS} = \mathbf{X}_{:, :, 0, :}$ \Comment{$\mathbf{X}_\text{CLS} \in \mathbb{R}^{B \times N \times d}$}
        \State $\mathbf{P} = \textbf{LayerNorm}(\textbf{GELU}( \mathbf{X}_\text{CLS}\mathbf{W}_1 + \mathbf{b}_1))$ \Comment{$\mathbf{P} \in \mathbb{R}^{B \times N \times d}$}
        
        \For{each pair of packets $(i, j)$}
            \State $\hat{z}_{ij} = \textbf{Softmax}( (\mathbf{P}_{:,i,:} - \mathbf{P}_{:,j,:})\mathbf{W}_2 + \mathbf{b}_2)$ \Comment{$\hat{z}_{ij} \in \mathbb{R}^{B \times 2}$}
        \EndFor
        
        \State $\mathcal{L}_{\text{PRPP}} = -\sum_{i,j,i \neq j} \mathbb{I}(i < j) \log(\hat{z}_{ij}) + \mathbb{I}(i > j) \log(1 - \hat{z}_{ij})$
        
        \State \textbf{Return} $\mathcal{L}_{\text{PRPP}}$
    \end{algorithmic}
    \label{alg:PRPP_task}
\end{algorithm}

\begin{algorithm}[htp]
    \caption{FCL Task}
    \begin{algorithmic}[1]
        \Require $\mathbf{X} \in \mathbb{R}^{B \times N \times L \times d}$ is the output from the MIETT Encoder, where $B$ is the batch size, $N$ is the number of packets, $L$ is the packet length, and $d$ is the dimension of embedding.
        \State $\mathbf{X}_\text{CLS} = \mathbf{X}_{:, :, 0, :}$ 
        \Comment{$\mathbf{X}_\text{CLS} \in \mathbb{R}^{B \times N \times d}$}
        \State $\mathbf{C} = \textbf{MLP}(\mathbf{X}_\text{CLS})$
        \Comment{$\mathbf{C} \in \mathbb{R}^{B \times N \times d}$}

        \For{each pair of packets $(i_1 j_1, i_2 j_2)$ where $i_1, i_2 \in [1, B]$ and $j_1, j_2 \in [1, N]$}
            \State $\mathbf{S}_{i_1 j_1, i_2 j_2} = \frac{\mathbf{C}_{i_1 j_1}^T \mathbf{C}_{i_2 j_2}}{\|\mathbf{C}_{i_1 j_1}\| \|\mathbf{C}_{i_2 j_2}\|}$ \Comment{Cosine similarity}
        \EndFor
        
        \State $\mathcal{L}_{\text{FCL}} = - \sum_{\substack{i_1, j_1, j_2 \\ j_1 \neq j_2}} \log\frac{\exp(\mathbf{S}_{i_1 j_1, i_1 j_2})}{\exp(\mathbf{S}_{i_1 j_1, i_1 j_2}) + \sum\limits_{i_2 \neq i_1} \exp(\mathbf{S}_{i_1 j_1, i_2 j_2})}$
        
        \State \textbf{Return} $\mathcal{L}_{\text{FCL}}$
    \end{algorithmic}
    \label{alg:FCL_task}
\end{algorithm}

\begin{algorithm}[htp]
    \caption{Fine-Tuning Task}
    \begin{algorithmic}[1]
        \Require $\mathbf{X} \in \mathbb{R}^{B \times N \times L \times d}$ is the output from the MIETT Encoder, where $B$ is the batch size, $N$ is the number of packets, $L$ is the packet length, and $d$ is the dimension of embedding.
        
        \State $\mathbf{X}_\text{CLS} = \mathbf{X}_{:, :, 0, :}$ 
        \Comment{$\mathbf{X}_\text{CLS} \in \mathbb{R}^{B \times N \times d}$}
        \State $\mathbf{X}_\text{mean} = \textbf{MeanPooling}(\mathbf{X}_\text{CLS})$ \Comment{$\mathbf{X}_\text{mean} \in \mathbb{R}^{B \times d}$}
        
        \State $\hat{y} = \textbf{MLP}(\mathbf{X}_\text{mean})$ \Comment{$\hat{y} \in \mathbb{R}^{B \times C}$, where $C$ is the number of classes}
        
        \State $\mathcal{L}_{\text{ft}} = - \sum_{c} y_c \log(\hat{y}_c)$
        
        \State \textbf{Return} $\mathcal{L}_{\text{ft}}$
    \end{algorithmic}
    \label{alg:finetuning_task}
\end{algorithm}

\subsection{Implementation Details.}
During the pre-training stage, we set the training steps to 150,000 and randomly select five of the first ten packets for training. The masking ratio for the Masked Flow Prediction (MFP) task is set to 15\%. The weights for the Packet Relative Position Prediction (PRPP) and Flow Contrastive Learning (FCL) tasks are both set to 0.2. In the fine-tuning stage, we train for 30 epochs using the first five packets.

For both stages, the packet length ($L$) is set to 128, the number of packets ($N$) is set to 5, the embedding dimension ($d$) is set to 768, and the number of Two-Level Attention (TLA) layers is set to 12. The learning rate is set to $2 \times 10^{-5}$, and the AdamW optimizer is used. All experiments are conducted on a server with two NVIDIA RTX A6000 GPUs. The PyTorch version is 2.3.0, and the random seed is fixed at 0 for reproducibility. Pre-processed hexadecimal data are provided by Netbench~\cite{netbench}, so we do not need to handle raw traffic data.

\end{document}